# Graphene-based technologies for energy applications, challenges and perspectives

Etienne Quesnel[1], Frédéric Roux[1], Fabrice Emieux[1], Pascal Faucherand[1], Emmanuel Kymakis[2], George Volonakis[3], Feliciano Giustino[3], Beatriz Martín-García[4,5], Iwan Moreels[4,5], Selmiye Alkan Gürsel[6], Ayşe Bayrakçeken Yurtcan[7], Vito Di Noto[8], Alexandr Talyzin[9], Igor Baburin[10], Diana Tranca[10], Gotthard Seifert[10], Luigi Crema[11], Giorgio Speranza[11], Valentina Tozzini[12], Paolo Bondavalli[13], Grégory Pognon[13], Cristina Botas[14], Daniel Carriazo[14,15], Gurpreet Singh[14], Teófilo Rojo[14], Gunwoo Kim[16,17], Wanjing Yu[16,17], Clare P Grey[16,17] and Vittorio Pellegrini[4]

[1] CEA, Liten, 17 rue des Martyrs, 38054 Grenoble Cedex 9, France

[2] Center of Materials Technology and Photonics & Electrical Engineering Department, Technological Educational Institute (TEI) of Crete,

Heraklion, 71004 Crete, Greece

[3] Department of Materials, University of Oxford, Parks Road, Oxford, OX1 3PH, UK

[4] Graphene Labs, Istituto Italiano di Tecnologia, Graphene Labs, Via Morego 30, 16163 Genova, Italy

[5] Nanochemistry Department, Istituto Italiano di Tecnologia,Via Morego 30, 16163 Genova, Italy

[6] Faculty of Engineering & Natural Sciences, Sabanci University, 34956 Istanbul, Turkey

[7] Department of Chemical Engineering, Atatürk University, 25240 Erzurum, Turkey

[8] Department of Chemical Sciences, Universitadegli Studi di Padova, Padova, Italy

[9] Department of Physics, Umeå University, Umeå, SE-901 87, Sweden

[10] Theoretical Chemistry, Technische Universität Dresden, Bergstr. 66b, 01062 Dresden, Germany

[11] Fondazione Bruno Kessler, REET Unit, Via alla Cascata 56/C, 38123 Trento, Italy

[12] Istituto Nanoscienze del Cnr, NEST Scuola Normale Superiore, Piazza San Silvestro 12, 56127 Pisa, Italy

[13] Thales Research and Technology, Palaiseau 91767, France

[14] CIC EnergiGUNE, Parque Tecnológico de Álava, Albert Einstein 48, 01510 Miñano, Álava, Spain

[15] IKERBASQUE, Basque Foundation for Science, Bilbao, Spain

[16] Department of Cambridge, University of Cambridge, Lensfield Road, Cambridge CB2 1EW, UK

17 Cambridge Graphene Centre, University of Cambridge, Cambridge CB3 0FA, UK

E-mail: etienne.quesnel@cea.fr

[16] Department of Cambridge, University of Cambridge, Lensfield Road, Cambridge CB2 1EW, UK

17 Cambridge Graphene Centre, University of Cambridge, Cambridge CB3 0FA, UK

E-mail: etienne.quesnel@cea.fr

**Abstract**

Here we report on technology developments implemented into the Graphene Flagship European project for the integration of graphene and graphene-related materials (GRMs) into energy application devices. Many of the technologies investigated so far aim at producing composite materials associating graphene or GRMs with either metal or semiconducting nanocrystals or other carbon nanostructures (e.g., CNT, graphite). These composites can be used favourably as hydrogen storage materials or solar cell absorbers. They can also provide better performing electrodes for fuel cells, batteries, or supercapacitors. For photovoltaic (PV) electrodes, where thin layers and interface engineering are required, surface technologies are preferred. We are using conventional vacuum processes to integrate graphene as well as radically new approaches based on laser irradiation strategies. For each application, the potential of implemented technologies is then presented on the basis of selected experimental and modelling results. It is shown in particular how some of these technologies can maximize the benefit taken from GRM integration. The technical challenges still to be addressed are highlighted and perspectives derived from the running works emphasized.

**Introduction**

Graphene and related two-dimensional (2D) materials constitute the material basis of one of the most promising and versatile enabling nanotechnologies, in particular for energy applications [1]. The 2D crystals combine high electrical conductivity and a huge surface-to-weight ratio, making them highly suitable for storing electrical charge, gas storing, and catalytic reactions. Various energy devices could benefit from these materials, i.e., batteries, supercapacitors, $H_2$ storage tanks, fuels cells, and photovoltaic (PV) cells. The interest in 2D materials for energy applications comes not only from their properties, but also from the possibility of producing and processing them in large quantities in a cost-effective manner. Printable inks, for example, are a gateway to the realization of new-generation electrodes in energy storage and conversion devices. The challenge ahead is to demonstrate that they indeed meet energy industry requirements. Another challenge is also to develop disruptive technologies in which 2D materials not only replace traditional electrodes, but, more importantly, enable whole new device concepts in the longer term.

Within the European 'Graphene Flagship' project [2], researchers are very active in this area. Their activities are mainly focused on the use of graphene and graphene-related materials (GRMs). Their first objective is to develop key enabling graphene-based integration technologies that can be efficiently introduced in the energy-device manufacturing value chain. This means assessing various technological approaches that are safe, environmentally friendly, and scalable in an industrial environment while remaining cost-effective. Following the science and technology roadmap for graphene recently published [3], another key objective is to evaluate the actual value added by these technologies with respect to current device performances. At Horizon 2020, the goal is to provide European industry with sustainable and competitive solutions. In this timeframe, applicative objectives were already defined in relation to building, portable, or transport applications. They deal, in particular, with a low-cost roll-to-roll PV technology demonstrating solar cells with more than 10% stable efficiency, advanced electrodes using GRMs, 2D material-based composites for lighter batteries (energy capacity target: 300 Wh kg$^{-1}$), lighter H$_2$ tanks (H$_2$ storage gravimetric target: 5.5 W%), and more sustainable fuel cells utilizing less platinum (target: 10 kW gPt$^{-1}$).

Supported by device modelling, our research approach combines careful selection of graphene raw material followed by dedicated graphene functionalization prior to integration into the energy device. These two technological steps are particularly crucial and require innovative processes for material engineering to create operating material architectures able to show a real improvement with respect to conventional device technologies. On the basis of recent investigations implemented within the energy work package of the Graphene Flagship project, this article reports on various integration examples. Our purpose is to discuss and assess the potential of graphene and GRMs once integrated into a device, keeping in mind that this integration can affect the initial properties of the pristine material. With this in mind, the technical challenges still to be addressed will be highlighted, but, more importantly, we will see how some graphene-based technologies developed for specific applications can maximize the benefit taken from various GRMs. Finally, perspectives derived from performing theoretical and experimental studies will be also emphasized.

**Photovoltaics**

To date, bulk silicon solar cells dominate the PV market with 91% market share [4]. This market is mainly dedicated to PV modules for large-scale stationary plants and building rooftop systems. This first-generation solar technology is fully mature, with the best solar-to-electric power conversion efficiency (PCE) up to 25% [5] at the laboratory scale, not so far from the theoretical limit (29%) [6]. With only 9% market share, the second generation of solar technology based on thin-film solar cells (TFSCs) is well behind in terms of market but also performances. While this technology has been introduced to decrease the cost of solar cells, the maximum PCE demonstrated today is only ~20%. On this very competitive market of domestic electricity, lower-cost solar technologies (<0.5 €/W) [7] are definitely required. To that end, the emerging PV technologies, i.e., organic photovoltaics (OPVs), perovskite solar cells (PeSCs), and quantum dot (QD) -based devices associated with roll-to-roll processing technologies are the most promising, provided that their performances (efficiency, durability) are drastically improved. Graphene and GRM integration can serve this objective. As a diversification of conventional PV module markets, mobile or indoor applications, smart fabrics, semi-transparent solar cells (smart windows, green house powering), and biomedical applications are emerging. These alternative markets could take advantage of thin-film and OPV technologies. The requirements to meet in these cases go beyond solar cell efficiency and include various additional key functionalities that graphene integration could help to achieve, like PV device conformity, flexibility, environmental compatibility (bio medium), or tuneable optical transparency or colour.

Ideally, a solar cell must exhibit maximum light absorption for optimum generation of photocarriers collected with minimum electrical losses. Meeting these requirements needs solar-cell building blocks specifically tailored for each PV technology. Device integration of graphene is precisely used in the Graphene Flagship project to design novel solar cell elements like (i) transparent and conductive (TC) electrodes for TFSCs and OPVs, (ii) tuneable adaptive buffer layers for OPVs, and (iii) a graphene-inorganic hybrid PV absorber or electrodes for QD and PeSC solar cells.

**Pure graphene TC electrodes**

A maximum optical transparency (T) to electrical resistance (R) ratio for the TC electrode (facing the sunlight) and a high internal quantum efficiency (IQE) are prerequisites to build solar cells. In that respect, single-layer graphene (SLG) electrodes have been shown to exhibit interesting properties. When integrated in a-Si:H solar cells they can outperform the conventional ZnO:Al electrodes by bringing higher IQE at a short wavelength (figure 1(a)). This is because of the superior transmittance of the graphene electrode in this spectral region (see insert, figure 1(a)). However, under full sun exposure (figure 1(b)), the photocurrent collection remains poor with graphene because of its excessive sheet resistance, one to two orders of magnitude higher than those of ZnO:Al. This example illustrates the potential of graphene electrodes for various applications. Indoor light harvesting with blue-rich light-emitting diode lighting or visible semi-transparent solar cells could be preferably considered. This underlines, however, the way to go to make graphene a PV material, since minimum requirements are typically T > 85% and R < 50$\Omega/\square$. At the research level, such material does exist [8]; however, it still remains a true challenge for industry. An alternative approach could be developing *in-situ* graphene doping strategies during PV cell manufacturing that could open the way to technical solutions easier to implement in industry.

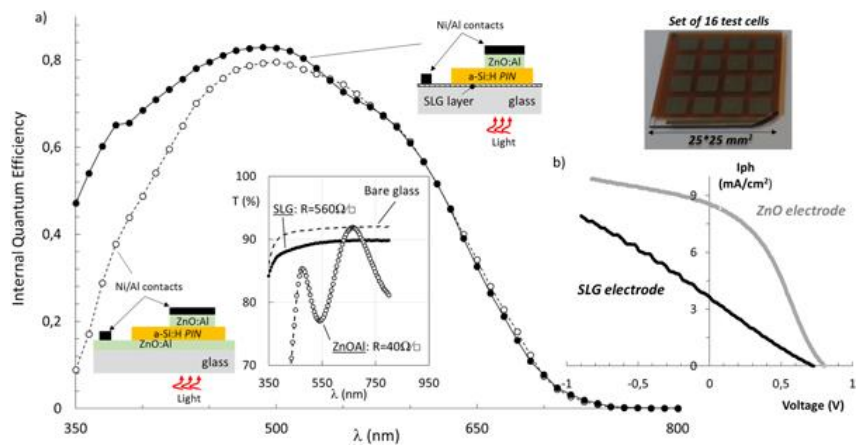

**Figure 1.** (a) IQE curve of a-Si:H PV cells with SLG/glass and ZnO:Al/glass TC electrodes. IQE($\lambda$) = EQE($\lambda$)/(1-R ($\lambda$)). EQE is the external quantum efficiency and R($\lambda$) the reflection measured on the TC electrode (light side). Insert: The transmittance curve of the TC electrode. (b) I/V curve under one sun exposure (standard AM1.5G spectrum, 100 mW cm$^{-2}$, 25 °C) for both kinds of electrodes measured on typical sets of test cells (picture).

**Graphene-based adaptive electronic buffer layers**

Hole and electron transport layers play a key role in bulk heterojunction OPV stacks by controlling the charge carrier extraction and transport towards the OPV devices electrodes, as well as by protecting sensitive photoactive blends from humidity and materials inter-diffusion. Graphene and its derivatives can be used to optimize this process in an OPV device. Since charge collection involves transport through different materials, the management of PV stack interfaces demands a fine tuning of the electronic work function (WF) of the respective layers for energy-band alignment control. An efficient modification of graphene WF was demonstrated [9] using a novel laser-based doping technology. Photochemical doping of chlorine atoms on the edge of graphene oxide (GO) flakes (figure 2(a)) can be effectively achieved by laser irradiation of pristine GO in the presence of a dopant chloride (Cl$_2$) precursor gas. The doping rate is simply controlled by adjusting the irradiation time, and the WF can be tuned, covering a wide energy range, from 4.9–5.23 eV (figure 2(b)). The resulting GO-Cl-based OPV devices (insert, figure 2(b)) demonstrate improved PV characteristics compared to the control devices, incorporating the current-state-of-the-art PEDOT:PSS hole transporter. Moreover, it improves the cell durability, a key issue towards OPV commercialization, by avoiding the use of a hygroscopic PEDOT:PSS layer. In the same manner, Li doping of GO layers [10] can be easily achieved to produce low-WF (4.3 eV) GO-Li layers and facilitate the electron extraction at the opposite top electrode (figure 2(c)). These examples show that GO inks associated with the laser technology and the functionalization process constitute a smart toolbox for much more efficient organic devices. It offers a unique and easy-to-use technology perfectly adapted to roll-to-roll solution processing, with a high potential in the OPV field and, more generally, in organic electronics. At medium term, a roll-to-roll compatible graphene-based OPV device reaching 10% efficiency will be demonstrated (figure 2(d)).

To validate the laser-based doping methodology and predict the full potential of this approach, density-functional theory (DFT) calculations were implemented. As an example, to probe the origin of the reported WF modification introduced by the Cl atoms, first-principles DFT calculations within the local-density approximation were performed for zig-zag and armchair graphene nanoribbons, both pristine and edge-functionalized with chlorine. Twenty-five percent of the edges of the model nanoribbons shown in figures 3(a) and (b) were functionalized. The structures were fully optimized, and it was found that the functionalization of the zig-zag edges is accompanied by an energy gain of 0.3 eV per Cl atom, while Cl atoms on armchair edges are only marginally stable. In order to analyze the WF change induced by the functional groups, we calculated the total charge density and the electrostatic potential around the pristine and the functionalized ribbons. The differences between these charge densities and potentials ($\Delta V$) are shown in figures 3(c) and (d), respectively. This $\Delta V$ is the key driver of the WF modification. We calculate WF shifts of 0.15 eV and 0.26 eV for zig-zag and armchair edges, respectively. These results qualitatively agree with the WF modification (0.25–0.3 eV) observed when functionalizing graphene-related structures with Cl atoms. These preliminary investigations have been extended to other functional attachments, as well as graphene flakes of various shapes and sizes. A comprehensive model of WF modification by covalent functionalization will be reported elsewhere.

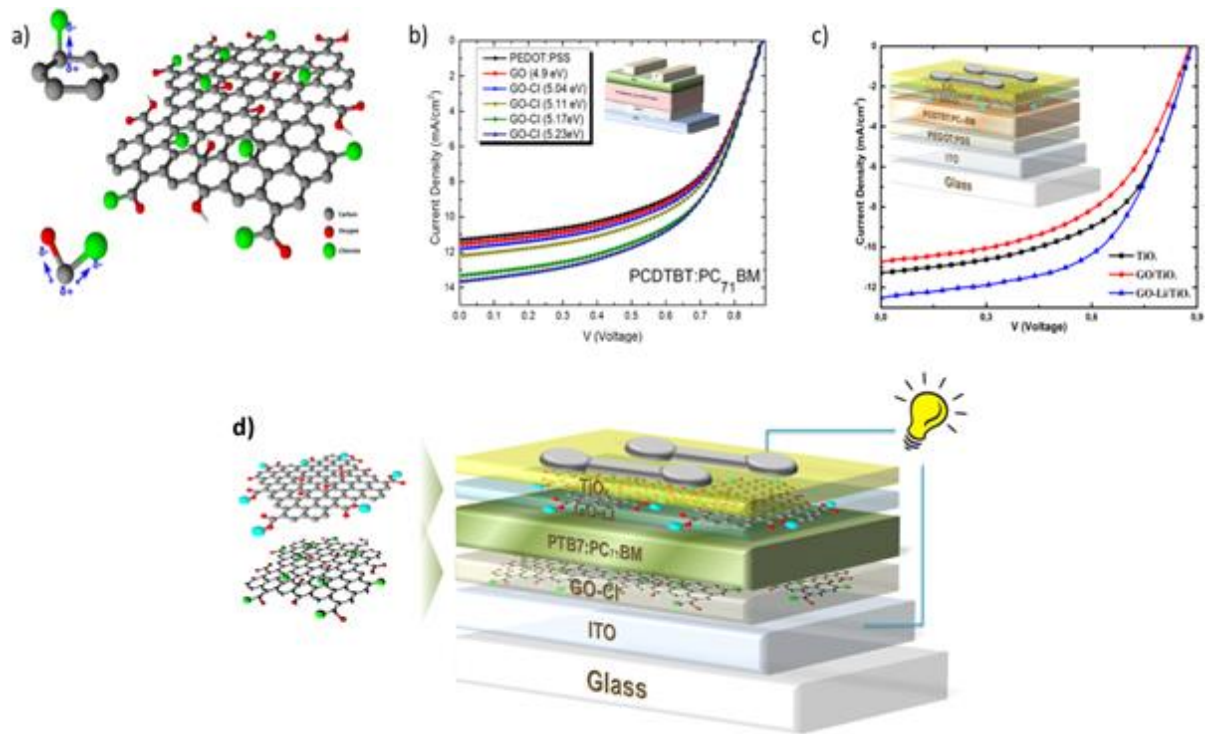

**Figure 2.** (a) Chlorine functionalization of GO; (b) Current density-voltage (J-V) curves under solar illumination (100 mW cm$^{-2}$) for various degrees of chlorine doping of GO-based hole transport layer deposited onto the bottom electrode; (c) J-V curves with and without adaptive Li-neutralized GO-based electron interfacial layer used near the top electrode; and (d) OPV device incorporating graphene-based (GO-Cl and GO-Li) buffer layers.

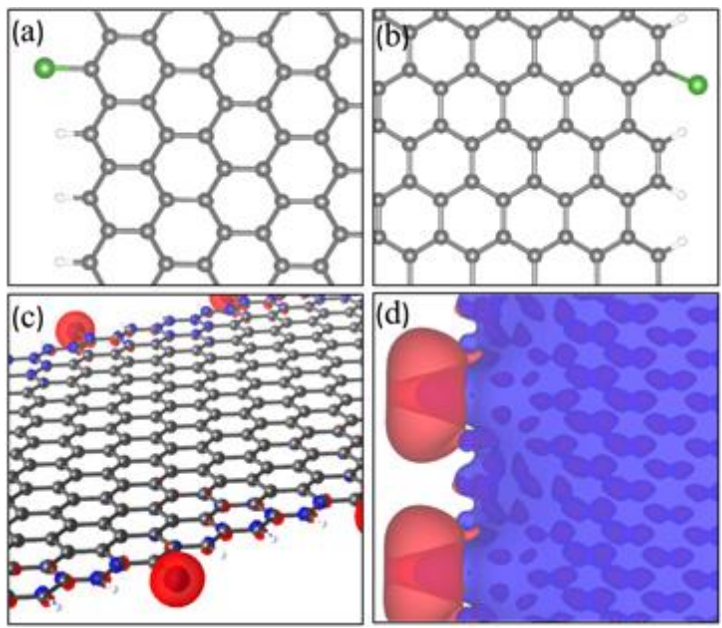

**Figure 3.** Model of Cl-functionalized graphene zig-zag (a) and armchair (b) nanoribbons. (c) Total charge density difference between Cl-functionalized and pristine zig-zag GNR. (d) Electrostatic potential difference induced by the Cl functionalization for a zig-zag GNR. (Charge density (potential) isosurface level: 0.05 e/a$_o^3$ (0.1 eV); positive isosurface values: blue, negative: red, C atoms: gray, Cl atoms: green, and H atoms: white.)

**Hybrid inorganic material/graphene compounds**

In theory, maximum PV conversion requires precise matching of the PV absorber bandgap to the solar spectrum, or tuning of spectra between the different absorbers in the case of a multi-junction solar cell [11]. Nanostructured PV cells based on semiconductor nanocrystals or QDs are a promising route towards the design of new absorbers showing various optical bandgaps, as these can be adjusted with high precision via the particle size [12]. In addition, the use of near infra-red absorbers such as colloidal PbS QDs allows harvesting a part of the solar spectrum that is not covered by silicon technology [13]. However, the introduction of nanostructured materials implies a tremendous increase of specific surface area (SSA) and led to a longstanding effort to improve the carrier mobility in QD close-packed films [14]. Here, incorporating graphene into the QD layer is expected to introduce channels to favour photocarrier transport.

Thus, by fabricating QD-graphene hybrid materials, the aim is to combine the optical properties of the QDs with the transport properties of 2D graphene. The challenge lies in achieving a large contact area and short distance between both materials to favour carrier transfer from the absorber to the charge-transport layer [15]. To maintain maximal flexibility for both constituents of the hybrid material, we decided on a solution-processable, covalent-linking approach based on the functionalization of reduced graphene oxide (rGO) with short-chained linker molecules, followed by anchoring QDs to the functionalized sheets (figure 4(a)). While graphene has larger carrier mobility, rGO facilitates the coupling of organic linkers [16, 17] via surface functional groups, thus providing a more suitable alternative for our method. In agreement with recent literature results [18, 19], we expect an efficient electronic coupling between the materials. Charge transfer from the QDs to the rGO is also supported by the PbS/rGO band alignment (figure 4(b)). More generally, the assembly via covalent linking provides a low-cost, solution-processable technology for QD/graphene-based hybrid devices. This strategy could apply to a large variety of QDs ($Cu_2S$, $CuInSe_2$, etc,) with potential applications for PVs and detectors [20, 21]. PV cell integration studies are ongoing to validate the viability of this approach.

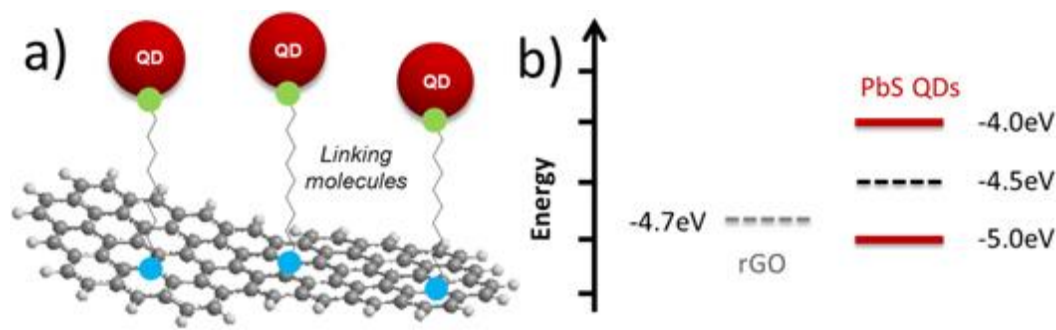

**Figure 4.** (a) Scheme of the QD/rGO coupling approach. (b) Approximate energy level diagram [12, 20, 22] supporting the concept of charge transfer between PbS QDs and rGO.

**Fuel cells**

Contrary to the PV industry, the fuel cell industry is still under construction, despite decades of research and developments. Beyond stationary applications and niche market applications (portable power sources), great expectations are foreseen for the car industry. Today, the massive introduction of exchange-membrane fuel cells on the car market is slowed down by various external factors, i.e., lack of light/high-capacity hydrogen storage tanks and nearly no refuelling stations. But the cost of the Pt catalyst used in such devices remains one of the main bottlenecks. Today, the typical platinum utilization efficiency is around 1 kW $gPt^{-1}$. This value must be increased by one order of magnitude at Horizon 2020 to comply with catalyst loading relevant to automotive applications. The durability of fuel cells is another serious concern, with the poisoning or aggregation of catalyst Pt particles that reduce the delivered electrical power. The typical lifetime requirement for proton exchange membrane fuel cells (PEMFCs) is >5000 h for transportation and >40 000 h for stationary applications.

Carbon-supported Pt catalysts are commonly used as anode and cathode catalysts for PEMFCs (figure 5(a)). The key point is to provide active metal nanoparticles with a high surface area. In that context, graphene is seen as a good means to develop novel catalyst supports, enabling better catalyst dispersion and anchoring to the graphene support. Both are expected to limit the catalyst aggregation during cell operation, in particular. Higher utilization efficiency, stability, and lifetime of the catalyst are thus expected with a perspective in the short term of reduced Pt content and in the longer term of no Pt at all.

Within the Flagship project, two approaches to enhance both performance and durability of fuel cells are considered. In the first approach, the ultimate goal is to reduce the amount of Pt used as a catalyst. In that case, we keep working with PEMFC devices, using a protonic membrane in an acidic medium (see figure 5(a)), and graphene- and rGO-based supports are used to control the Pt catalyst dispersion. The second approach is quite different, since a non-platinoid catalyst and anion exchange membrane fuel cells (AEMFCs) are considered. This radically new route based on CN-covered graphene supports should lead to much cheaper fuel cells.

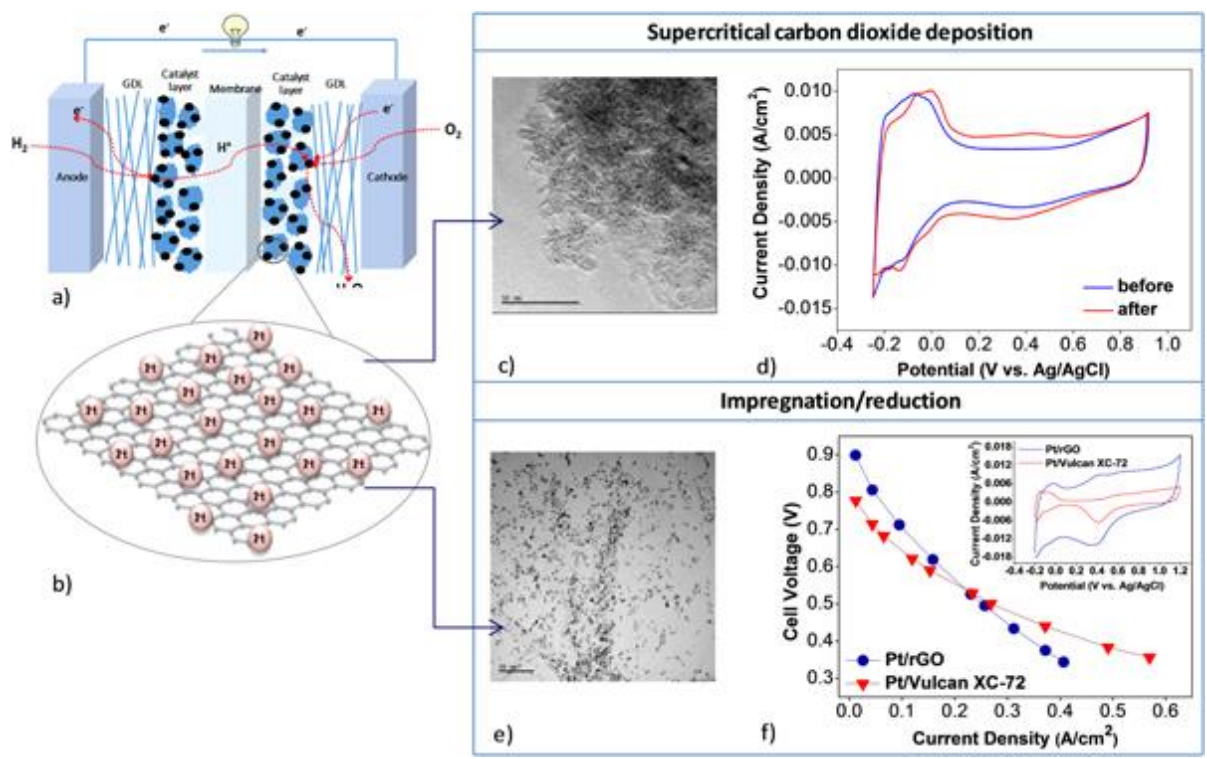

**Figure 5.** (a) PEMFC architecture; (b) Pt/GNP or Pt/rGO; (c) TEM image of Pt/GNP; (d) Cyclic voltammograms of Pt/GNP before and after carbon corrosion test; (e) TEM image of Pt/rGO; and

(f) PEMFC performance of Pt/rGO and Pt/Vulcan XC-72 at 65 °C and full humidification. Inset: Cyclic voltammograms of Pt/rGO ad Pt/Vulcan XC-72.

**Graphene nanoplatelets and rGO supports**

The electrocatalytic activity of fuel cell catalysts strongly depends on several parameters, such as catalyst support, catalyst preparation technique, catalyst precursor, accessibility of the metal catalyst, and fuel cell testing conditions [23]. In use, the higher the number of active nanoparticles in the catalyst, the better the fuel cell performance is. The catalyst preparation techniques aim thus to obtain small and highly dispersed nanoparticles over the support material.

In the first approach, graphene nanoplatelets (GNPs) or rGO have been employed as catalyst support in the gas-diffusion electrodes instead of commercial carbon black. Graphene-supported Pt nanoparticles (Pt/GNP or Pt/rGO, figure 5(b)) were developed by means of both the supercritical carbon dioxide ($scCO_2$) deposition method and the impregnation/reduction method. The first technique is gaining increasing attention because of being environmentally friendly and leaving no residue in the catalyst. It involves the dissolution of the Pt organometallic precursor in an $scCO_2$ environment and adsorption of the dissolved precursor on the carbon support material. The adsorbed Pt precursor can be converted into its metallic form with thermal reduction in a nitrogen atmosphere. The transmission electron microscopy (TEM) image (figure 5(c)) revealed highly dispersed spherical and small-size Pt nanoparticles over the GNP support. Figure 5(d) shows the cyclic voltammogram (CV) of the Pt/GNP catalyst before and after a carbon corrosion test applied for 24 h. This test is used as an accelerated degradation test and consists in applying 1.2 V for a given time. On this example, the loading of the Pt/GNP catalyst is low (5.7%), and the electrochemically active surface area (ECSA) measured before the corrosion test is high (ECSA: 90 $m^2\ g^{-1}$). After the corrosion test, about 18% ECSA loss was observed; this is an acceptable value, despite the high amount of graphene in the catalyst. In the same manner, the preparation of highly dispersed and uniformly decorated Pt nanoparticles with a small particle size (4–6 nm) on rGO can be achieved by the impregnation method, using ethylene glycol reflux (figure 5(e)). Fuel cell electrodes are simply prepared by spraying of a catalyst ink on the carbon paper. As shown in the inset of figure 3(f), synthesized Pt/rGO exhibits significantly higher electrocatalytic activity than commercial Vulcan XC-72-supported Pt nanoparticles (Pt/Vulcan XC-72) [24]. Moreover, in PEMFC test conditions, Pt/rGO exhibits higher open-circuit voltage and superior performance at low current densities compared to Pt/Vulcan XC-72 with the same Pt content (0.25 mg $cm^{-2}$) [21]. This may be attributed to the better dispersion and small particle size of the Pt/rGO catalyst. Considerable enhancements in electrocatalytic activities, stability, and fuel cell performance have been thus achieved for graphene-supported Pt nanoparticles as compared with carbon black-supported nanoparticles. These preliminary results are promising to achieve our ultimate goal of reducing Pt content in the electrodes (target: 10 kW $gPt^{-1}$).

## CN-coated graphene supports

An alternative and extremely innovative approach to devise next-generation oxygen reduction reaction (ORR) electrocatalysts (ECs) for application in low-temperature fuel cells (e.g., PEMFCs and AEMFCs) consists in covering graphene and GRMs with a thin carbon nitride (CN) layer, which embeds the active sites; thus, $M_1 \cdots M_n$-CN/GRM ECs are obtained. The GRMs template the EC morphology, acting as a 'core'; the CN layer, indicated as the 'shell,' stabilizes the active sites in 'nitrogen coordination nests,' improving ORR performance and durability in operating conditions. $M_1 \cdots M_n$-CN/GRM ECs are synthesized with a very flexible preparation protocol, which proved capable to: (a) modulate the chemical composition of the active sites, to boost the intrinsic ORR turnover frequency; (b) control the concentration and distribution of nitrogen in the shell; and (c) achieve a high dispersion of active sites, thus facilitating the transport of reactants and products in the system [25].

Early studies were aimed at addressing the specific issues raised by the adoption of GRMs as the core in ECs obtained with the proposed preparation protocol. The synthesis of the first $M_1 \cdots M_n$-CN/GRM ECs built on the experience gained on other core systems [26, 27] and yielded products exhibiting very interesting features and a promising ORR performance. As a first proof of concept, here we discuss the PtNi-CN$_l$ 900/graphene ORR EC, which is labelled in accordance with the conventions described elsewhere [23].

The morphology of PtNi-CN$_l$ 900/graphene is characterized by a very disordered stacking of graphene sheets covered by a porous CN shell (e.g., see the lower portion of figure 6(a)). The latter embeds well-dispersed PtNi$_x$ nanoparticles. The semi-quantitative decomposition of powder x-ray patterns (see figure 6(b)) matches perfectly this picture, highlighting: (a) PtNi$_x$ alloy nanoparticles with the typical fcc structure and exhibiting a grain size of ∼10 nm; and (b) a very small graphene (002) peak, which witnesses the almost complete exfoliation of the graphene core support. The successful introduction of a CN shell in PtNi-CN$_l$ 900/graphene is also clear from HR-TGA results collected in an oxidizing atmosphere (see figure 6(c)). The CN shell (comprising ∼14 wt% of the EC) undergoes oxidative degradation at ∼530 °C, while the decomposition of the graphene core occurs at ∼680 °C. It is pointed out that the conventional active carbon support of the Pt/C reference EC is degraded at ∼430 °C, witnessing a much lower tolerance to an oxidizing environment (such as that at the cathode compartment of an operating fuel cell). This evidence is a first hint of an improved durability of the proposed CN-based core–shell ORR ECs. In an acid environment, PtNi-CN$_l$ 900/graphene already shows an ORR performance, selectivity, and reaction mechanism very close to those characterizing the Pt/C reference (see figures 6(d) and (e)). Further efforts are currently underway to optimize the chemical composition and improve the dispersion of the ORR active sites to push the performance and durability of the $M_1 \cdots M_n$-CN/GRM ECs decisively beyond the state of the art.

Finally, it is evidenced that, owing to its outstanding flexibility, the proposed preparation protocol is also suitable to devise ORR ECs that do not include platinum-group metals, with a dramatic drop in cost; preliminary results in an alkaline environment (data not shown) are very promising.

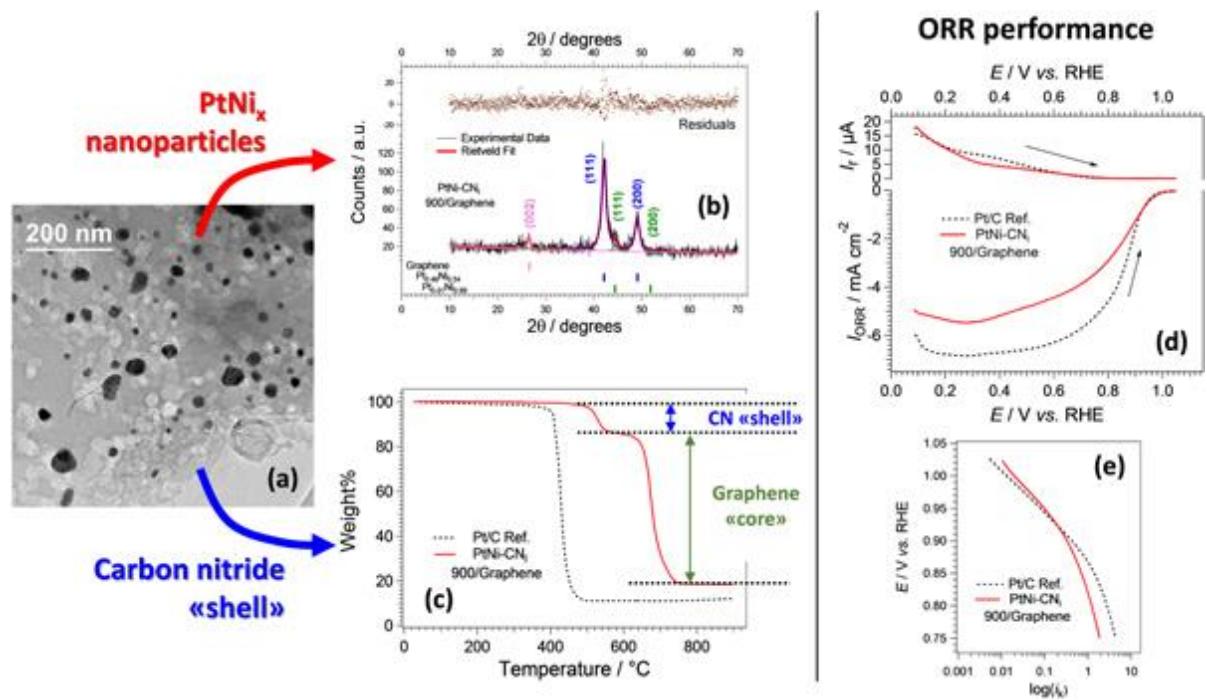

**Figure 6.** Early development of ORR $M_1\cdots M_n$-CN/GRM ECs: the case of PtNi-CN$_l$ 900/Graphene. (a) Morphology investigation by HR-TEM; (b) semi-quantitative phase identification and analysis on powder XRD spectra; (c) HR-TGA profiles in an oxidizing environment; (d) ORR performance studied by CV-TF-RRDE method in an acid environment; and (e) Tafel plot in the ORR.

**H$_2$ storage**

Hydrogen storage is a key issue for the development of hydrogen-driven fuel-cell electric vehicles and stationary applications. The current solution to store hydrogen on board of vehicles is to use liquefaction or high-pressure gas tanks. However, using pressures of 700 Bar creates safety issues, while use of cryo-tanks does not meet car industry requirements. According to the US Department of Energy the target H2 storage system should be capable of a gravimetric capacity of 5.5 W% and a volumetric capacity of 40 kg m$^{-3}$, both at 293 K. For higher-capacity storage solutions, alternative technologies based on hydrogen storage using a solid compounded powder of metal hydrides are emerging. With Mg-based hydrides, the capacity of storage at ambient temperature is relatively high (a few %wt. [28]), but gas release is energy-consuming and requires heating at 300–400 °C. In the same manner, graphene and GRMs are expected to be good candidates as materials for H$_2$ storage because of their high SSAs.

Graphene can be interesting for hydrogen storage applications in several ways: for physisorption of molecular hydrogen, for chemisorption of hydrogen, and as an addictive to various composite structures, e.g., as support for light metal hydride nanoparticles. The Flagship project is aimed to examine all three of these research directions.

**Very high SSA GRMs for H$_2$ physisorption**

This H$_2$ storage mechanism was carefully assessed by systematic hydrogen uptake measurements on various carbon-based materials with a wide range of SSA values. Clear hydrogen adsorption trends were confidently established for rGO and activated rGO samples for ambient and 77 K temperatures (figure 7(a) [29]). These data provide important reference points for development of GRMs with improved hydrogen uptakes. They show a precise correlation of H$_2$ uptake with surface area for all studied samples. That is why the possibility to increase the surface area of graphene beyond 2650 m$^2$ g$^{-1}$ (the limit for pristine defect-free graphene) is also investigated, using theoretical modelling. It is shown [30] that incorporation of defects, preferably small 'holes' up to 10 Å in diameter, results in the increase of theoretically possible surface areas approaching ∼5000 m$^2$ g$^{-1}$ (see figure 7(b)). This leads to promising hydrogen storage capacities of up to 6.5 wt.% at 77 K (15 bar), as estimated from classical grand canonical Monte Carlo simulations. The main challenge of this approach is the synthesis of material in a large quantity, preserving a well-defined graphene interlayer separation in the prepared samples. The best of the so-far experimentally prepared samples (activated rGO) demonstrated BET surface areas of ∼2900 m$^2$ g$^{-1}$ and hydrogen storage capacity of ∼5.5 wt.% at 77 K and ∼0.9 wt.% at 290 K (120 bar H$_2$).

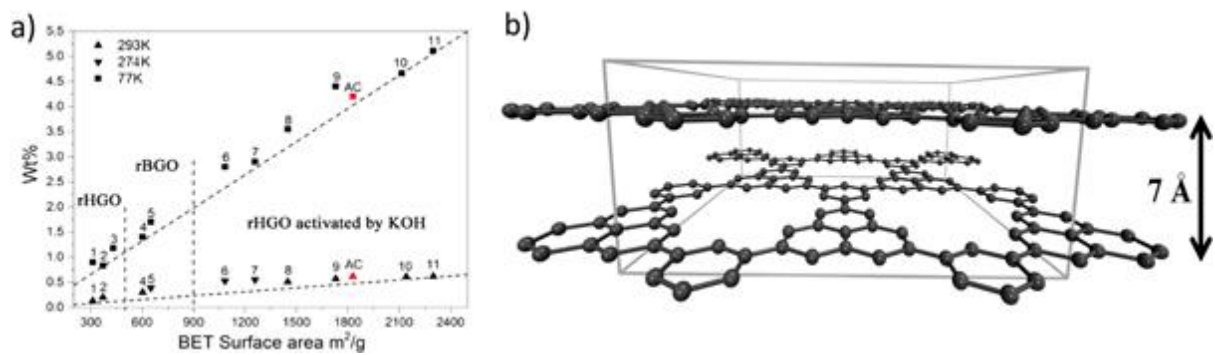

**Figure 7.** (a) Hydrogen uptakes by graphene samples measured by volumetric method at ambient (120 Bar) and 77 K (50 Bar) temperatures and plotted vs surface area. (b) 3D stacking of perforated graphene layers with surface area of 5100 m$^2$ g$^{-1}$ (adapted from reference [30]).

**Decoration of graphene with metal atoms for H$_2$ chemisorption**

The storage capacity of graphene can be increased by surface functionalization. A promising route is decoration with alkaline-earth or transition metals. A transition metal will be covalently bound to carbon. However, due to the high cohesive energy, transition metals tend to cluster on the surface of carbon nanostructures. Elements with smaller cohesive energies, like alkaline-earth metals, are more suitable as dopants to carbon nanostructures [31]. In recent years, graphene nanoribbons (GNRs) have been intensively studied due to their electronic properties. The GNR properties are strongly dependent on the atomic structures and the chemistry of the edges [32]. For example, GNRs can be passivated by adsorption of oxygen along their edges, leading to oxygenated graphene nanoribbons (GNR-O). Subsequently, alkaline-earth metals can be chemisorbed at the edges of such GNR-O structures. For example, for Ca adsorbed on the oxygen edge of an armchair-oxygenated graphene nanoribbon (AGNR-O), the binding energy is 5.41 eV/Ca compared to 0.74 eV/Ca on top of the carbon atoms. DFT calculations demonstrated the binding of up to four hydrogen molecules to the alkaline-earth metal in such M-functionalized AGNR-O (M = Ca/Mg) structures (see figure 8). The adsorption energy is increased by a factor of five, from 0.05 eV/$H_2$ for 'naked' nanoribbons (AGNR) to 0.250 eV/$H_2$ for Mg-functionalized AGNR-O. This is a very promising result concerning the increase of the hydrogen storage capacity of graphene by functionalization.

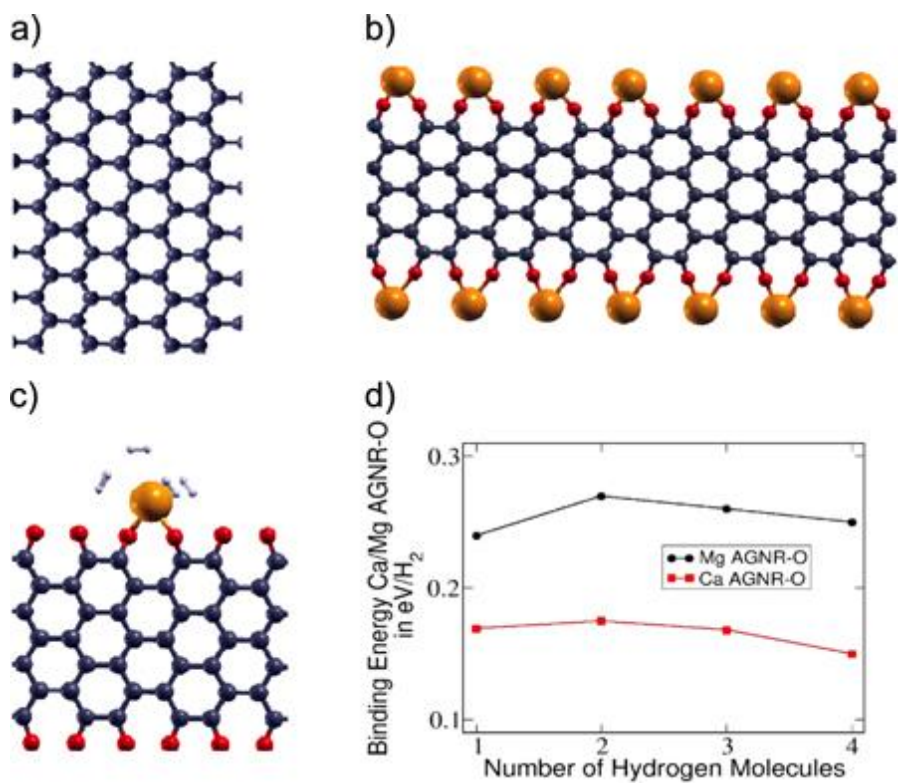

**Figure 8.** Optimized structures of (a) graphene, (b) oxygen-passivated graphene nanoribbon (AGNR-O), (c) adsorbed $H_2$ molecules, and (d) calculated binding energies.

**GRM-hydride composites for $H_2$ chemisorption**

Magnesium hydride (MgH$_2$) has generated a lot of interest as a potential hydrogen storage material because it has a high storage capacity (7.6 wt.%), availability, and safety [33]. However, the sluggish hydrogen-sorption kinetics and high operating temperatures for hydrogen absorption/desorption hinder its practical application [34]. Extensive research has been made to solve those problems, mainly by particle-size reduction through mechanical milling and catalyst additives, but the improvements were not significant enough to break the limitations and reach substantial gravimetric and volumetric densities at moderate enthalpies of reaction [35, 36]. Recently, both theoretical- and experimental- method studies have reported that more significant improvements are possible but only with much smaller MgH$_2$ nanoclusters (<5 nm) [37–41]. Calculations demonstrate that for very small (MgH$_2$)n clusters (n < 19), the enthalpy of decomposition sharply reduces with cluster size [42]. For a Mg$_9$H$_{18}$ cluster of approximately 0.9 nm diameter, a desorption enthalpy of 63 kJ mol$^{-1}$ was calculated [43], from which a decomposition temperature of about 200 °C can be estimated. Although several successful syntheses were reported, there are still problems in terms of yield, scalability, and overall performance of the materials, including hydrogen capacity and stability [35, 44, 45]. Such small-size nanoparticles are easily prone to aggregation and sintering, especially when the hydride particles undergo phase change during hydrogenation/dehydrogenation. In this respect, the nano-confinement method implemented in the Flagship project could offer a promising alternative for synthesizing stable small-Mg nanoparticles with a technology potentially scalable for mass production. As a first proof of concept, multiwall carbon nanotubes (CNTs) were utilized as a matrix to induce growth of Mg particles within the CNT walls, thus limiting the Mg cluster size. The synthesis of MWCNT-MgH$_2$ hybrid material was carried out through a catalytic hydrogenation of Mg anthracene complex (MgA.3THF). A THF suspension of MgA.3THF is mixed with MWCNTs and, in proper catalytic reaction conditions (CrCl$_3$), the formation of MgH$_2$ starts immediately after transfer of the mixture into a pressure reactor vessel for hydrogenation under 30 bar H$_2$ at 60 °C and during 15 h. TEM imaging of the resulting nanoparticles (NPs) demonstrates that the size of the MgH$_2$ clusters ranges from 1–5 nm, as desired (figure 9(a)). Actually, larger NPs (20–50 nm) were also observed that grew on the MWCNT's surface. As a consequence, a remarkable H$_2$ desorption of hybrid material is observed at 80° C and reaches a maximum at ∼167 °C, testifying to the beneficial role of the Mg cluster-size confinement approach (figures 9(b) and (c)). However, system optimization is still required, since further H$_2$ desorption occurs at higher temperatures (336 °C and 363 °C), related to the presence of larger MgH$_2$ NPs behaving as bulk magnesium hydride. The use of a GRM-based confinement matrix is under investigation to solve this issue (ongoing experiments).

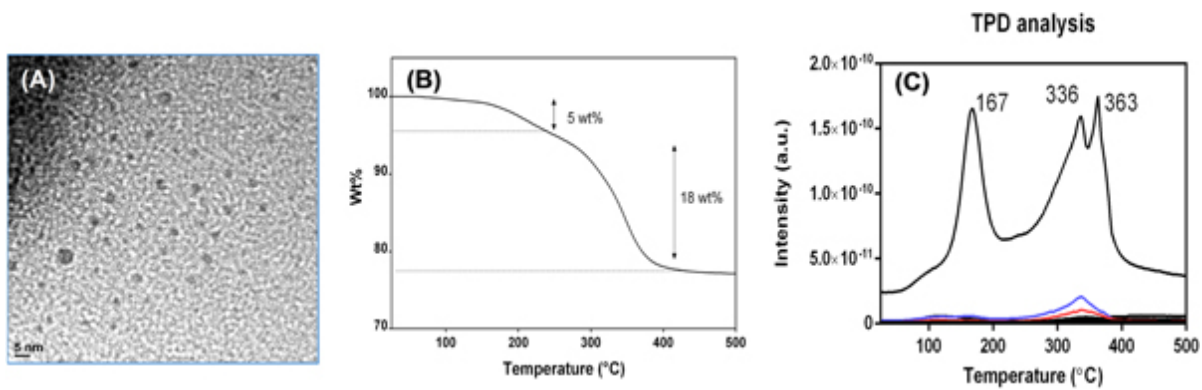

**Figure 9.** (a) TEM imaging of Mg NPs detached from infiltrated MWCNT: average dimension is in the 1–5 nm range; (b) thermogravimetric analysis (TGA), and (c) temperature-programmed desorption (TPD) spectra of the Mg-infiltrated MWCNTs.

**Mechanically actuable GRM-based H$_2$ hosting materials**

Graphene has exceptional mechanical properties, which imply the possibility of rippling it, statically or dynamically. This peculiar property allows innovative and unique strategies to store hydrogen. The modelling activities implemented in the Flagship project already demonstrated various mechanisms particularly interesting for H$_2$ storage or transport (figure 10). We showed that curvature-enhanced reactivity allows us to selectively control chemi(de) sorption of hydrogen [46, 47] (figure 10(a)) and that convexities and defects can be used to control graphene decoration with gas-adsorption enhancer metals such as Ti [48]. Moreover, recent simulation results also predicted that the dynamical modulation of curvature produced by flexural phonons of nanoscale wavelength could be used also to transport hydrogen (or other gases) through graphene multilayers over macroscopic scales [49] (figure 10(b)). This effect could be used in graphene-based devices to move or pump gas, eventually also increasing GD within 3D scaffolds. Our efforts are currently devoted to exploring efficient strategies (mechanical, optical, or electromagnetic) to control curvature, statically or dynamically [50]. This approach could give rise to novel controllable H$_2$ storage tanks.

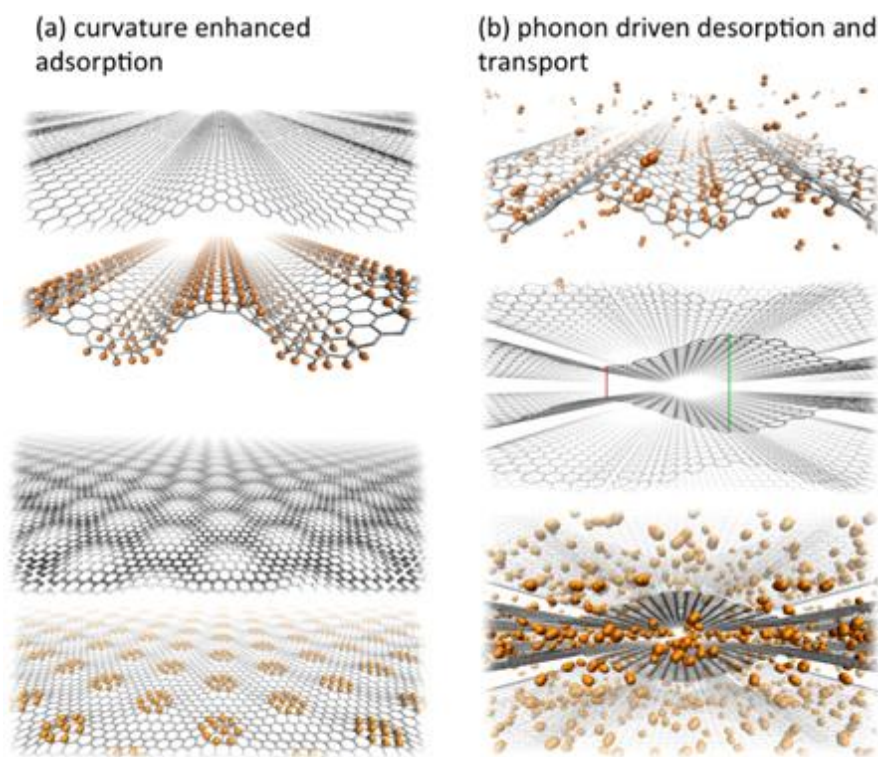

**Figure 10.** Curvature-controlled storage and transport. (a) Hydrogen chemisorption or decoration with other chemicals is enhanced over convexities. (b) Hydrogen is released upon curvature inversion (e.g., produced by a traveling flexural phonon). If phonons are excited in multilayers, the resulting traveling cavities can actively transport the gas.

**Electrochemical storage**

For decades, rechargeable batteries and supercapacitors have been the basic electrochemical devices used to store electricity. Their performance depends on both their capacity to store electrical charges (electrons, ions) as measured by their gravimetric energy density (Wh/kg) or to store/release this energy in a minimum of time, namely, their power density (kW/kg). Because of their high SSA and stable electrochemical and thermal properties, graphene and GRMs have been widely investigated to manufacture higher-capacity storage devices. As shown in a very recent review paper [1], there are actually many ways of integrating GRMs in these devices, using more or less sophisticated technological routes. In this very competitive international context, the Flagship researchers are concentrating their efforts on the development of safe and easy-to-handle material processes. While these technologies are expected to enable safe, clean, and cost-effective manufacturing of storage device electrodes, in some cases they can even foster more efficient storage mechanisms.

**CNT/Graphene composite for supercapacitors**

Supercapacitors are electrochemical devices that combine the high-energy storage capability of batteries with the high-power delivery capability of capacitors [51, 52]. Electrochemical double-layer capacitors (EDLCs) have been developed to provide power pulses for a wide range of applications such as hybrid electric vehicles, the electric utility industry (e.g., emergency backup power, grid system stability improvement), and consumer electronics. The first example of EDLCs based on CNTs was published by Niu *et al* [53] in 1997. CNTs are based on graphene sheets (one or more) rolled up to form a seamless tube of variable diameter, ranging from 1 ~ 100 nm. The recent availability of graphene and GRMs has pushed scientists to explore the potentialities of GRMs for EDLC fabrication. Graphene nano-sheets, when used alone, are likely to stack and agglomerate, reducing the available active surface area. In return, CNTs and graphene nano-sheets, when mixed together, are expected to prevent both restacking of graphene and formation of CNT bundles, thus enhancing the active surface area of the mix and giving rise to rapid diffusion pathways for the electrolyte ions. CNTs also serve as a binder to hold the graphene nano-sheets together, preventing disintegration of the electrode structure into the electrolyte.

In the Graphene Flagship project, we study the properties of GO- or graphene-CNT mixtures as a function of relative composition (%) and weight for the fabrication of supercapacitor electrodes. The electrode fabrication is performed using a novel dynamic deposition process set-up at Thales Research and Technology [54] (figure 11). This equipment allows high control of deposited material thickness and uniformity. Compared to the conventional filtration method (bucky paper), it could constitute a real breakthrough, because there is no limitation in electrode areas (up to tenths of $dm^2$), material thickness (nms to $\mu$ms), and substrate nature.

The first fabricated electrodes were using inks of graphite and CNTs (50%) in NMP solvent (N-Methyl-Pyrrolidone). The graphite was exfoliated by two sonication steps of 18 h (alone and with CNTs mixed together). These samples show specific capacitance and power densities of around 25 F $g^{-1}$ and 16 kW $Kg^{-1}$, respectively, for 2.5 mg material [55]. The same work done with mixtures of graphene and CNTs (50%) provides much higher performances with capacitance and power densities of 120 F $g^{-1}$ and 92.6 kW $Kg^{-1}$, respectively. The noticeable power enhancement clearly indicates that the larger conductance of graphene dramatically reduces the electrode resistance. The high capacitance density measured with an aqueous electrolyte ($LiNO_3$ 3M) is related to the high SSAs of the exfoliated graphene provided by the IIT partner (~2000 $m^2$ $g^{-1}$). Despite the simplicity of spraying technology, the specific capacitance obtained so far is comparable to the EDLC state of the art but higher in terms of power density [56]. In an attempt of developing a safer technology by replacing the toxic NMP solvent with water, graphene flakes and CNTs were replaced by GO flakes and oxidized CNTs. After spray deposition and post-annealing at 300 °C for some hours, capacitance densities similar to G-CNT-based devices were actually measured (120 F $g^{-1}$). However, the power density remains under 25 kW $Kg^{-1}$. This technology development is still ongoing. It already enables the exploitation of graphene-based nanomaterial for supercapacitors with a safe and easily scalable process.

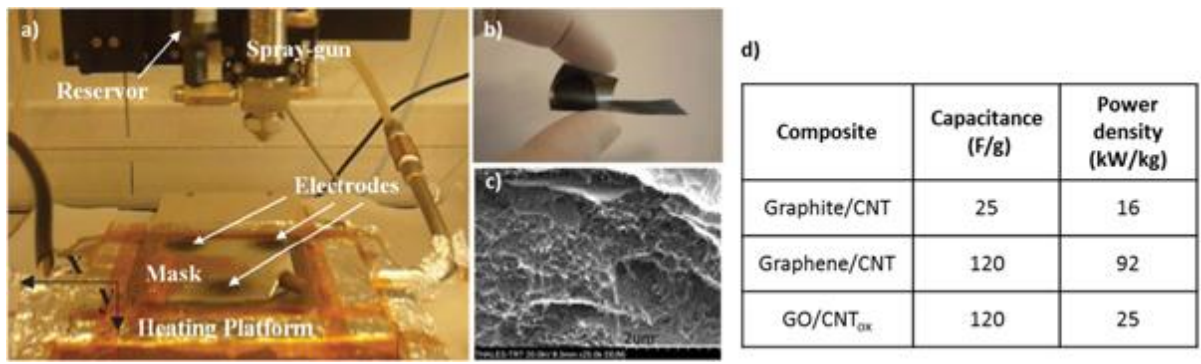

**Figure 11.** (a) Dynamic spray-gun deposition set-up used for the deposition of mixtures of nanomaterial graphite collectors; (b) flexible electrode deposited on a graphite collector; (c) SEM cross section of a graphite collector covered with a mixture of sprayed graphene/graphite (50%)/CNTs (50%); (d) typical electrode performance versus composite material.

**Nanocrystal/graphene composite anode for batteries**

Graphene or GO flakes intermixed with a Li ion inorganic host (Sn, Si…) is an established investigation route to increase the anode capacity while keeping the electrode mechanically stable and electronically conducting. The challenge is to accommodate the large volume changes undergone by the host material after lithiation and after hundreds of insertion/extraction cycles. Within the Graphene Flagship project we are developing new strategies to obtain composites formed by Sn and $SnO_2$ particles wrapped by graphene, which provides a matrix support that buffers the volume changes that occur during the cycling of these materials. Among them, we are exploring the mechanical mixture of Sn NPs together with GO and its subsequent thermal reduction under inert atmosphere. As an alternative approach, an *in-situ* synthesis, where the precursor of the metal (Sn salt) is directly mixed with the GO suspension and then thermally reduced to induce the synthesis of a rather homogeneous distribution of metal particles, was also evaluated. This later synthetic pathway not only promoted the growth of sub-micron particles that are embedded within the rGO layers but also allowed us to process the electrodes as macro-porous aerogels or self-standing films. These prepared sponges or films could be directly assembled into the cell without the need of using any metallic support or its further processing together with binders and additional conductive additives (figure 12).

The electrochemical behaviour over cycling of the *in-situ* prepared $SnO_2$@rGO samples showed remarkably high capacity values compared with the mechanically mixed materials and with the Sn-free samples previously prepared in our lab, showing in the former case stable reversible capacities at 50 mA of ∼1000 mAh per gram of electrode after 150 cycles [57]. Figure 13 shows the charge-discharge curves registered in the 2nd cycle of an electrode formed by macro-porous rGO and by an aerogel formed by the SnO$_2$@rGO composite that evidence the improvement attained when Sn-based particles were included into the material, compared to the Sn-free electrodes.

Even though SnO$_2$@rGO shows high-capacity values and significant stability in terms of lithium ion cycleability (figure 13(b)), it is important to mention here that the major challenge in developing the full cell based on rGO incorporation in the anode will be to overcome the irreversible capacity that occurs in graphene-based anode materials due to the high SSA of the graphene. This is highly dependent on the graphene content and hence the solid-electrolyte interphase (SEI) formation in the composite anodes.

To get an idea of the full potential of such SnO$_2$@rGO electrodes, another concern is to understand why the measured specific capacities exceed the theoretical bulk capacity of SnO$_2$ (782 mAh g$^{-1}$). The concept of nanosize Li ion hosting material assumes that lithiation is similar to that of the bulk material but with a better ability to accommodate the mechanical stress induced by this lithiation. In reality, the detailed electrochemical reaction processes and mechanism for Li storage in such materials are unclear and may be different from the bulk, as, for example, these materials will be less pulverized during Li insertion and extraction. As a model system for studying Li storage mechanism by *ex-situ* NMR spectroscopy, SnO$_2$ NP-pillared rGO composites were prepared with high electrochemical activity via a facile preparation method. rGO and colloidal SnCl$_4$ particles were mixed, and a subsequent reaction with hydrazine resulted in the formation of SnO$_2$ nanocrystals on the rGO surface. The formation of a crystalline SnO$_2$ phase within the rGO was confirmed by powder x-ray diffraction, the electrode comprising rGO and SnO$_2$ nanocrystals with a typical size of 5 nm (figure 14(a)).

Unlike graphite, which reacts with Li via an insertion mechanism, bulk SnO$_2$ reacts initially via a conversion reaction (to form Li$_2$O and Sn; theoretical capacity 782 mAh g$^{-1}$), Sn then reacting via an alloying reaction, resulting in an overall theoretical capacity of 1493 mAh g$^{-1}$. With such a system, a reversible specific capacity over 1000 mAh g$^{-1}$ is measured after 30 cycles (figure 14(b)) a stable reversible specific capacity of 973 mAh g$^{-1}$ is reached after 100 cycles, suggesting that Li$_x$Sn alloys must be forming at the end of discharge. Preliminary $^{119}$Sn *ex-situ* NMR studies at various stages of (dis)charge of the SnO$_2$ NP-pillared rGO system show that SnO$_2$ is recovered after one cycle (figure 15). The formation of Li$_2$O is confirmed via $^7$Li NMR, consistent with the conversion reaction The $^{119}$Sn and $^7$Li NMR spectra differ from those observed for reactions of bulk Sn alloys, where distinct resonances due to the Li$_x$Sn alloy phases are seen. Instead, the $^{119}$Sn resonances shift and broaden (figure 15(b)) as the SnO$_2$ is reduced. The broad resonance is tentatively assigned to an alloy-phase Li$_x$Sn with a particle size that is sufficiently small to result in metallic behaviour. The results are consistent with both a conversion and alloying reaction mechanism for SnO$_2$, which gives rise to the high observed practical capacity.

This favourable Li storage mechanism is observed for composites containing 1–5 nm $SnO_2$ particles. Furthermore, the *in-situ* synthesis of Li hosting nanocrystals within the graphene-based scaffold results in significantly improved performance over both larger Sn particles and composites prepared by simply mixing the two components. NMR studies suggest different mechanisms for bulk versus nanosize particles.

This result should open a wide field of investigation for other nanostructured battery anodes, in particular for the preparation of Si@rGO composites. Silicon is known indeed for its very high theoretical Li insertion capacity (3572 mAh g$^{-1}$). The Si@rGO materials are difficult to prepare, since Si particles tend to agglomerate in larger secondary particles, but the present approach should help overcome this issue.

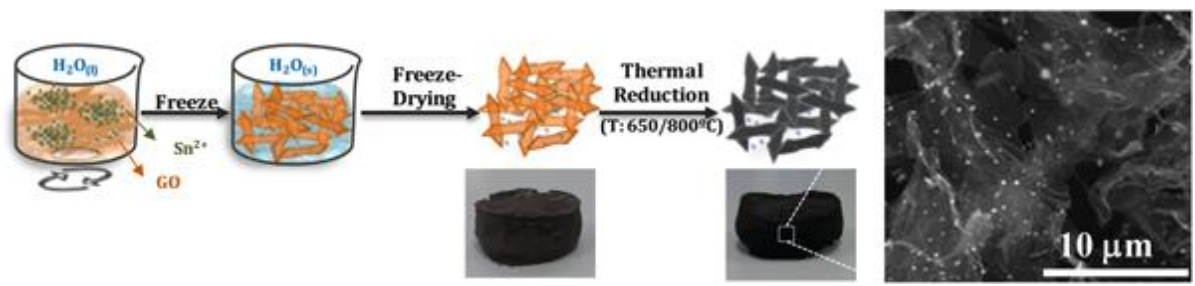

**Figure 12.** Schematic for the fabrication of $SnO_x$@rGO composite aerogels, which includes digital photographs of $SnO_2$@rGO sponges and the SEM image of this sample showing the opened macro-porous microstructure of this material.

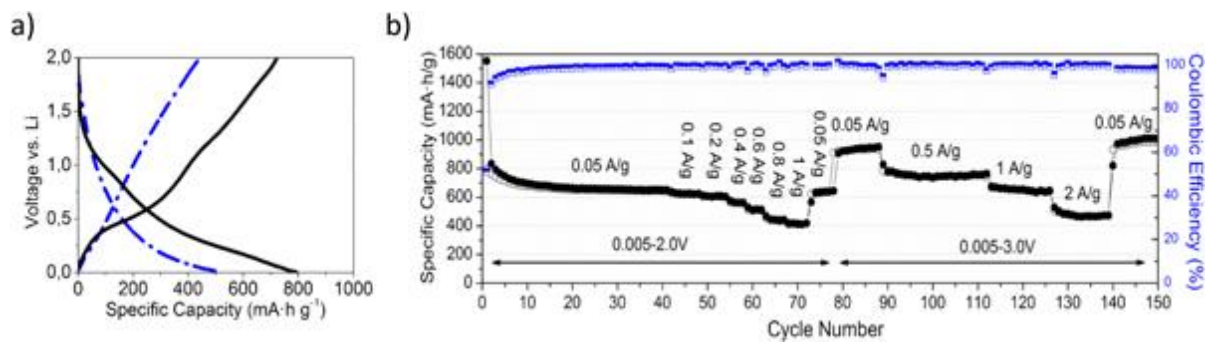

**Figure 13.** (a) Galvanostatic charge and discharge curves of aerogels formed by rGO (blue line) and the $SnO_2$@rGO (black line) in the 2nd cycle. (b) Specific capacity values per mass of electrode versus cycle at indicated charge-discharge rates of a $SnO_2$@rGO composite foam [from reference [57]].

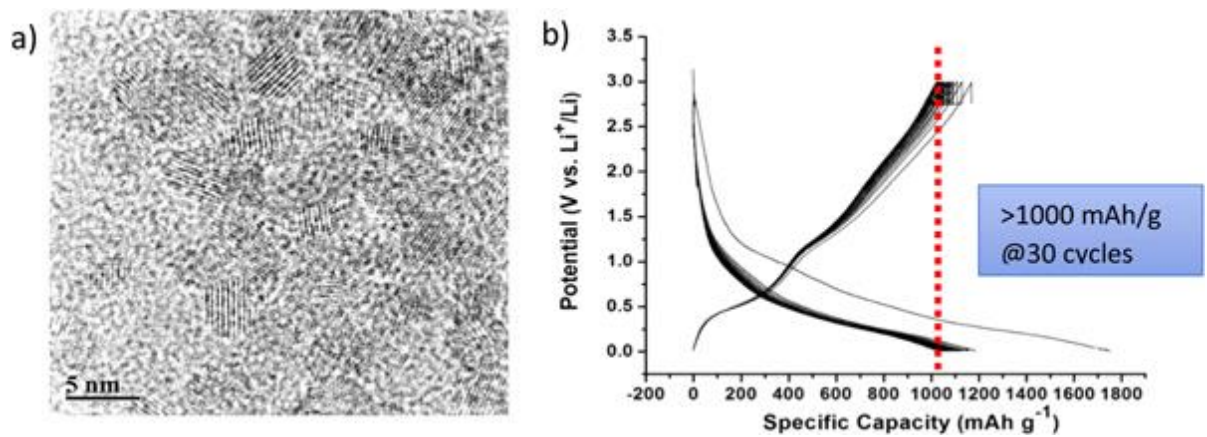

**Figure 14.** (a) TEM image of SnO$_2$ NP-pillared rGO composite, and (b) the electrochemical profiles of SnO$_2$ NP-pillared rGO composite for the first 30 charge and discharge cycles. The sample was cycled galvanostatically against Li metal (voltage cutoff between 3–0.005 V) with a current rate of 50 mA g$^{-1}$.

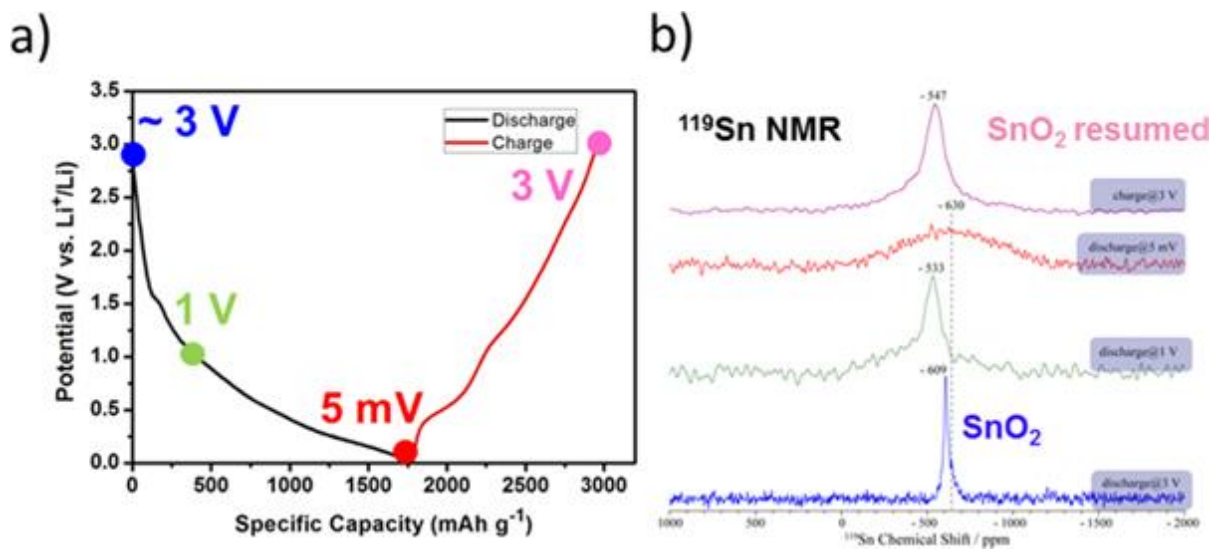

**Figure 15.** (a) The voltage profile of SnO$_2$ NP-pillared rGO composite during charge and discharge, and (b) solid-state $^{119}$Sn MAS NMR spectra at different states of charge, acquired at 16.4 T, with a MAS frequency of 60 kHz. The samples collected for *ex-situ* NMR are indicated on the graph.

**Conclusion and perspectives**

On the basis of material integration examples previously presented, several technologies and integration strategies seem to emerge as potentially promising routes for future graphene introduction into the energy device industry.

A first striking result concerns the generic use of graphene and GRMs (rGO) as hosting scaffolds to control the growth of functional nanosize (typically <5 nm) crystals starting from liquid precursors. The particular interest of such an approach is highlighted in different topics: in fuel-cell electrodes for Pt catalyst decoration, in hydrogen storage for the synthesis of improved metal hydride materials, or in Li-ion nanocrystal-based anodes ($SnO_2$). In all these cases, graphene can bring unexpected functionalities like well-dispersed nanometer-size Pt catalyst particles for higher catalytic activity in fuel cells, much lower $H_2$ desorption temperatures for hydrogen storage application, or more efficient Li ion storage mechanisms for battery anodes. Graphene appears, then, as a key enabling material that provides a clear added value to the final product. However, introducing such a high-SSA material like graphene can also bring serious drawbacks, like high irreversible capacity in battery electrodes, for instance. In the same manner, nanosizing of the hosting material can induce undesired electrode instability because of possible NP re-aggregation. Beyond device proof of concepts, these developments require, then, further material engineering to finely control the surface chemistry of the internal cavities where nanocrystal synthesis takes place inside the GRM scaffolds. In particular, it is essential to precisely control their internal surface energy to drive the nanocrystal nucleation and growth process during electrode decoration and/or to enhance the adhesion of grown nanocrystals and prevent re-aggregation. In the same manner, controlling the formation of the SEI in batteries, which competes with reversible lithium intercalation, is a fundamental problem that graphene integration could help understand and maybe control. These developments raise several fundamental material science questions that need to be more deeply addressed within the Graphene Flagship project to secure graphene integration into future energy devices.

Another very interesting result previously emphasized deals with the possibility to control and tune the surface chemistry and electronic properties of rGO surfaces using a laser-based doping technology. The fact that the electronic properties experimentally measured are confirmed by theory reinforces the reliability of this process. This easy-to-handle laser process is clearly a key enabling technology for future organic PVs and, more generally, organic electronics, including a very wide range of sensors. This very promising technological route will be strongly pushed further.

Last but not least, taking advantage of the high SSA of graphene to build efficient supercapacitors or hydrogen storage materials seems to be indeed a promising route. With respect to conventional activated carbon that already show noticeable SSAs (up to 2000 $m^2\ g^{-1}$), GRMs could in theory provide much higher SSA material architectures. Graphene-based material showing extremely high SSA (5000 $m^2\ g^{-1}$) is indeed theoretically possible by incorporating holes into the material. Our best result experimentally demonstrated so far is approaching 3000 $m^2\ g^{-1}$, not far from best state of the art [58]. However, engineering of such material to end up with a stable structure showing even higher SSA remains extremely challenging. In this regard, the dynamic spray technology now under development could be a good way to reinforce the mechanical stability of such compounds by building hierarchically ordered composite materials by a multi-layering process.

Finally, although graphene is regularly presented as the ideal material to replace transparent and conductive oxides (e.g., ZnO, ITO, $SnO_2$, etc), we confirm that its integration in PV devices actually requires further work to meet the very demanding PV specifications (T > 85% and R < 50$\Omega/\square$). In this respect, material and fundamental sciences are also necessary to identify and assess the better way to dope CVD graphene and to evaluate in which extent this doping can be controlled because of the relatively high initial contamination of metal-catalyzed CVD graphene sheets (metallic impurity concentration of typically ~$10^{12}$ at/$cm^2$ and more). An integrated approach, combining synthesis, characterization, theory, device fabrication, and scale-up is essential for progress in this field.

**Acknowledgments**

We acknowledge funding from the Graphene Flagship (contract no. CNECT-ICT-604391).